
\magnification1200
\font\bigbf=cmb10 scaled\magstep2
\def\uqg{{\cal U}_q(g)}
\line{\hfil December 93}
\bigskip
\line{\hfil \bigbf Universal R-matrix of reductive Lie algebras \hfil}
\bigskip
\line{A. Kundu \footnote{$^\dagger$}{Permanent Address:
 Saha Institute of Nuclear Physics,AF/1 Bidhan Nagar,Calcutta 700 064 India.}
\hfil }
\line{\it Fachbereich 17--Mathematik/ Informatik,
GH--Universit\"at Kassel \hfil}
\line{\it Holl\"andische Str. 36, 34109 Kassel, Germany \hfil }
\medskip
\line{  P. Truini \hfil}
\line{\it Dipartimento di Fisica, Universit\`a di Genova e I.N.F.N.\hfil}
\line{\it v. Dodecaneso, 33, 16146 Genova, Italy\hfil}
\vskip1.5truecm
\noindent
{\bf Abstract.} {\it A universal ${\cal R}$-matrix intertwining between the
coproduct structures  related to the universal deformations
of reductive Lie algebras is found in the explicit form using
twisting.}
\vskip1.5truecm
The universal $\cal R$-matrix is an abstraction of the quantum
$R$-matrix used
in the theory of integrable models. In the study of Hopf algebras it
becomes an object of its own interest and plays a central role
in the definition of quasitriangularity [1].
The explicit construction of the universal $\cal R$-matrix is a difficult
task in general, but following the prescription given by
Drinfeld [2] such an object has been found for deformations
of semisimple [3-5] as well as affine [6] Lie algebras.
In [6-7] the uniqueness of such solutions within the given ansatz
has also been shown. In the recent past a multiparameter deformation of
all reductive Lie algebras has been formulated [8] with the property
of universality in a certain class. Our aim is to build a universal
$\cal R$-matrix for this deformation and show that it is indeed a
quasitriangular Hopf algebra.
In order to do so we exploit the twisting method [9]
for introducing new parameters as well as for making the transition to
the reductive case. The physical motivation behind this construction
is to go back to the theory of integrable models again and use such
$\cal R$-matrix for building the associated quantum $R$-matrix and Lax
operators with spectral as well as color parameters. The color
parameters should be provided by the eigenvalues of the central
generators of the reductive Lie algebra in a given representation.
However in this letter we present only the explicit form of the
universal $\cal R$-matrix of reductive Lie algebras and postpone its
application to a forthcoming paper.
\medskip
Let $  g  $ be a reductive Lie algebra of rank $N$. Namely $g$ is the
direct sum of, say, $M$ simple Lie algebras plus an abelian center.
$H_1 , H_2, \ldots , H_N $ is the basis
of the Cartan algebra  of which $ H_i (1 \leq i\leq N_1) $
span the semisimple part and the remaining $N-N_1$ number of
 $H_{\alpha} , (N_1 < \alpha \leq N)$ belong to the center of the algebra.
Let $ {\bf a}_k$ be vectors with $N$
 components $a_{\ell k}$, such that  $a_{\alpha k} = 0$ for all $k$ and
  $ (N_1 < \alpha \leq N)$  , while  $a_{ij}= 2(\alpha _i \cdot\alpha _j)/
(\alpha _i \cdot \alpha _i)$ with $ (1 \leq i, j \leq N_1)
  $ is the Cartan matrix of the semisimple part. Let $X_i^{\pm}$
be the generators related to the simple roots $\alpha_{i}$ .
   It is established that
[8] for such algebras   the universal deformation $\uqg$
 can be defined such that
each simple component remains the same as that of the standard
one-parameter quantization
$$
[X_i^+,X_j^-]= \delta_{ij} { q_{ii}^{H_i\over 2} -
 q_{ii}^{ -{H_i\over 2}} \over
   q_{ii}^{1\over 2} - q_{ii}^{ - {1\over 2}}  } \equiv  \delta_{ij}
   [ H_i ]_{q_{ii}}
\eqno(1) $$
and $ f({\bf H}) X_i^{\pm} =  X_i^{\pm}  f({\bf H}+{\bf a}_i) $
with the Serre relation
$$
\sum_{0 \leq k \leq n}  (-1)^k
  \pmatrix{ n \cr k}_{q_{ii}} \left(  X_i^{\pm}  \right)^{n-k}
     X_j^{\pm} \left(   X_i^{\pm}  \right)^k = 0 ,
\eqno(2) $$
with $\ n= 1-a_{ij}
 $ and the notation $q_{ii}=e^{h_{\rho(i)}(\alpha_i\cdot
\alpha_t)}  $, where $(1 \leq  \rho(i) \leq
M)$ counts the number of simple components. Since $\rho$ is constant
on each simple component, we may use also the notation
$q_{\rho} = e^{h_{\rho (i)}}$.
The remaining deformation parameters on the other hand can be relegated to the
coalgebra structure defining the corresponding coproducts as
$$
 \Delta ( X_i^{\pm}) =  X_i^{\pm} \otimes   \Lambda _i^{\pm}
 +( \Lambda _i^{\pm})^{-1} \otimes    X_i^{\pm}
\eqno(3) $$
$$
 \Delta ( H_i) =  H_i \otimes {\bf I} +   {\bf I}  \otimes   H_i ,
\eqno(4) $$
where $\Lambda _i^{\pm} $, containing the parameters $v_{i\alpha}, t_{ik}$
$(t_{ik} = - t_{ki})$, has the form
$$
\Lambda _i^{\pm} \equiv q_{ii}^{  H_i\over 4} e^{\pm  { 1\over 2}
  \left( \sum_{j, k } t_{ik}  a^{kj} H_j  +  \sum_{\alpha}  v_{i \alpha}
    H_{\alpha} \right)}
\eqno(5) $$
The antipode ($S$) and co-unit ($\epsilon$) are given by
$$
 S( X_i^{\pm}) =-  q_{ii}^{\pm {1\over 2}}
   X_i^{\pm}, \ \
  S( H_i) =-  H_{i}
\ \   \hbox{and }  \ \ \epsilon (X_i^{\pm}) = \epsilon (H_i)=0.
\eqno(6) $$
This defines the universal deformation as a Hopf algebra.
\medskip
Our aim here is to find a universal
 ${\cal R}$-matrix ($\cal R$ is in a completion of $\uqg\otimes\uqg$ [1])
corresponding
to the above Hopf algebra , which would intertwine between the
 coproduct  $\Delta$ (3-5) and its permuted form  $\tilde \Delta
 = \sigma \Delta $ as
$$
{\cal R}   \Delta (a) =  \tilde \Delta (a) {\cal R}  \qquad \forall
a\in \uqg .
\eqno(7) $$
and satisfy the properties
$$
       (\Delta\otimes{\bf I}) {\cal R}  =   {\cal R}_{13}
         {\cal R}_{23} , \quad
          ({\bf I}\otimes\Delta) {\cal R}  =   {\cal R}_{13}
         {\cal R}_{12} .
\eqno(8) $$
Such a $\cal R$ would naturally endow the above Hopf algebra
with quasitriangularity.
\medskip
Following the argument of ref. [3-6] we construct first the
universal ${\cal R}$-matrix
for the case $t_{ik}=v_{i\alpha}=0$, i.e.
for the semisimple but untwisted
deformed Lie algebra depending only on the deformation parameters $q_\rho$
($1\leq \rho \leq M$). Denoting it by ${\cal R}_0$, we have
$$
 {\cal R}_0 = {\check R} K ,
\eqno(9) $$
where $K$ is expressed in terms of the Cartan generators only:
$$
 K = \exp \left(\sum_{ij}
 h_{\rho(i)} {(\alpha_i \cdot \alpha_i)\over 2}
{(\alpha_j \cdot \alpha_j)\over 2}     d^{ij}    H_i {\otimes}H_j \right),
\eqno(10) $$
with $    d^{ij}=  ( d^{-1})_{ij} $, where $d_{ij}= (\alpha_i
 \cdot \alpha_j )$ is the symmetrized Cartan matrix.
$\check R$ is then given in a factorized form as:
$$
     \check R  = \prod_{\rho=1}^M (  \check R^{(\rho)}), \ \ \hbox {with}
      \ \     \check R^{(\rho)} =   \prod_{\gamma \in \Delta_+^\rho
      } (  \check R^{(\rho)}_\gamma)
\eqno(11) $$
where $\Delta_+^\rho$ is the set of {\it all} positive roots belonging
to the $\rho$-th simple component with the prescribed normal ordering [6].
We have in turn
$$
\check R^{(\rho)}_\gamma = \exp_{q^{-1}_{\gamma \gamma}}
\left( a_\gamma^{-1}( q_\rho - q_\rho^{-1})( e_\gamma \otimes
e_{-\gamma}) \right)
\eqno(12) $$
where $ \  \exp_{q_{\gamma \gamma}} \ $ is the $q$-exponential function and
$a_\gamma$ is defined by the commutation relation among the following
elements of the {\it q-deformed} Cartan-Weyl basis [6]
$$
[e_{\gamma},e_{-\gamma}]= a_\gamma { q_{\gamma\gamma}^{h_\gamma /2} -
 q_{\gamma\gamma}^{ - {h_\gamma /2}} \over
   q_{\rho} - q_{\rho}^{ - 1}  }
\eqno(13) $$
In the above expression (13), if the index $\gamma$ correspods to the
non-simple root $\alpha_\gamma = \sum \alpha_i$ for certain simple
roots $\alpha_i$, then $h_\gamma = \sum h_i$ and
$$ h_i = {(\alpha _i \cdot \alpha _i) \over 2} H_i $$
$$ e_i = \left[ {(\alpha _i \cdot \alpha _i)\over 2}\right] _{q^2_\rho} ^{1/2}
\;   q_{ii}^{-1/4} X^\pm_i q_{ii}^{\mp H_i/4} $$

The  ${\cal R}_0(q_{ii})$-matrix thus obtained
complies with the coproduct of the form (3-4), where the
 $\Lambda_i^{\pm}$ operators contain only  $q_{ii}$  as deforming
  parameters:
$$       {\cal R}_0  \Delta_0 (a) =  \tilde \Delta_0 (a) {\cal R}_0
\qquad a\in \uqg \eqno(14) $$
where
$$
 \Delta_0 (   X_i^{\pm}) =
          X_i^{\pm} \otimes
            q_{ii}^{ H_i\over 4}
            +    q_{ii}^{- {H_i\over 4}}  \otimes
              X_i^{\pm}
\eqno(15) $$
\medskip
Next we include  the remaining parameters $t_{ij}$ and $ v_{i\alpha}$
to obtain
     ${\cal R}(q_{ii}, t_{ij}, v_{i\alpha})$
by implementing the twisting transformation [9] and
suitably choosing the twisting operators for the reductive case.
\medskip
We thus abtain the following universal $\cal R$-matrix associated to the
universal deformation of any reductive Lie algebra
$$
                    {\cal R }(q_{ii}, t_{ij}, v_{i\alpha})=
                     {\cal G}^{-1}(t_{ij}, v_{i\alpha})
           {\cal R }_0 (q_{ii})   {\cal G}^{-1}(t_{ij}, v_{i\alpha})
\eqno(16) $$
where ${\cal R}_0$ is given in explicit form through (9-12) and the twisting
operator $\cal G$ is given by
$$\eqalign{
        {\cal G}(t_{ij}, v_{i\alpha}) &=
\exp \left\{ {1\over 2} \left( \sum_{j \alpha}
      v_{j\alpha}   a^{jk} (  H_k {\otimes}H_\alpha -
H_\alpha {\otimes}H_k ) \right.\right.  \cr &+
    \left.\left. \sum_{ijkr}
      a^{ki}   t_{kr}  a^{rj}   H_i {\otimes}H_j \right)\right\}
\qquad (a^{ij} = (a^{-1})_{ij}) \cr }
\eqno(17) $$
\par
Note that  the  property    ${\cal G}_{21}^{-1}=
   {\cal G}_{12}$  which is essential for constructing a consistent
$\cal R$-matrix through eqn. (16),
 is satisfied due to the antisymmetry of $t_{kr}$ and
irrespective of the fact that $v_{j\alpha}$ is not antisymmetric
(recall that $ \{ v_{j\alpha} \}$
   may have $(N-N_1)\times N_1 $ number of independent elements).
For showing that the operator (17) may be taken as a twisting
operator , following Reshetikhin [9] $\cal G$ has to satisfy also
$$
       (\Delta\otimes{\bf I}) {\cal G}  =   {\cal G}_{13}
         {\cal G}_{23} , \quad
          ({\bf I}\otimes\Delta) {\cal G}  =   {\cal G}_{13}
         {\cal G}_{12} ,
\eqno(18) $$
along with
$$
         {\cal G}_{12}  {\cal G}_{13}
         {\cal G}_{23}
           =    {\cal G}_{23} {\cal G}_{13}
         {\cal G}_{12}
\eqno(19) $$
which however can be checked easily.
\medskip
We may see now that the coproduct is changed under such twisting as
$$
     \Delta ( a)= {\cal G} \
 \Delta_0 ( a) \ {\cal G}^{-1} \qquad a\in \uqg
\eqno(20) $$
yielding in particular
$$\eqalign{
    \Delta ( X_i^{\pm})&= {\cal G} \
 \Delta_0 ( X_i^{\pm}) \ {\cal G}^{-1}
  \cr &=  X_i^{\pm} \otimes
 q_{ii}^{H_i \over 4} e^{\pm  { 1\over 2}
  \left( \sum_{j, k } t_{ik}  a^{kj} H_j
   +  \sum_{\alpha } v_{i\alpha}  H_\alpha
     \right)}
 +  \cr &
  q_{ii}^{- { H_i\over 4}} e^{\mp  { 1\over 2}
  \left( \sum_{j, k } t_{ik}
    a^{kj} H_j + \sum_{\alpha } v_{i\alpha } H_\alpha
     \right)}
  \otimes    X_i^{\pm} \cr }
\eqno(21) $$
  and  $   \Delta ( H_i)= \Delta_0 ( H_i) $
and this coincides exactly with (3-5)
with all nontrivial parameters $ \ q_{ii}, \  t_{ij} $ and $ v_{i\alpha}$.
\par
Consequently, the universal ${\cal R}$-matrix thus obtained has the properties
(7-8) and this concludes the proof that the algebra introduced in [8]
is (pseudo [1]) quasitriangular. The total number of independent
parameters contained in $\cal R$ is $M + {1\over 2} N_1(N_1-1) + N_1(N-N_1)$
for a reductive Lie algebra of rank $N$ with $M$ simple components
plus a $(N-N_1)$-dimensional center.
\medskip
For the application of this construction to the theory of quantum
integrable systems we may consider different finite-dimensional
representations of the ${\cal R} $-matrix yielding
the corresponding braid group representations (BGR), color BGR  along
with the related $L^{\pm} $ matrices used in the
Faddeev-Reshetikhin-Takhtajan algebra (FRT) [10]. Moreover if it is
possible to formulate a suitable Yang-Baxterization scheme for
the FRT-algebra we may be able to construct genuine spectral
parameter dependent Lax operators and quantum $R$-matrices generating
thus new classes of quantum integrable models, beside the known
Toda field models on lattice.
This project has been successfully carried through for $gl(n)$ and
the result will be reported in a forthcoming paper [11].
\medskip
{\it Acknowledgement}
One of the authors (AK) thanks the Alexander von Humboldt foundation
and the I.N.F.N. (Italy) for the financial support.
\bigskip\noindent
\line{\bf References \hfil}
\medskip
\item{1.} Drinfeld V.G., {\it Quantum Groups} ( ICM Proceedings, Berkeley)
          798, 1986
\item{2.} Drinfeld V.G., {\it Algebra Anal.} {\bf 1} 30 (1989)
\item{3.} Rosso M., {\it Comm. Math. Phys.} {\bf 124} 307 (1989)
\item{4.} Kirillov A.N. and Reshetikhin N. {\it Comm. Math. Phys.} {\bf 134}
          421 (1990)
\item{5.} Levendorsky S.Z. and Soibelman Ya.S., RGU preprint (1990)
\item{6.} Khoroshkin S.M. and Tolstoy V.N.,
          {\it Comm. Math. Phys.} {\bf 141} 599 (1991)
\item{7.} Khoroshkin S.M. and Tolstoy V.N., Wroclaw preprint ITP UWr
          800/92
\item{8.} Truini P. and Varadarajan V.S., {\it Lett. Math. Phys.}
          {\bf 26} 53 (1992)
\item{}   Truini P. and Varadarajan V.S.,
          {\it Rev. Math. Phys.} {\bf 5} 363 (1993)
\item{9.} Reshetikhin N., {\it Lett. Math. Phys.} {\bf 20} 331 (1990)
\item{10.} Faddeev L., Reshetikhin N. and Takhtajan L.,
           in {\it Braid Groups, Knot Theory and Statistical Mechanics ,}
           eds. C.N.Yang  and M.L.Lie (World Scientific ) 97, 1989
\item{} Reshetikhin N. Yu., Takhtajan L.A. and Faddeev L.D. {\it Algebra
        and analysis } {\bf 1} 178 (1989)
\item{11.} Kundu A., Truini P., in preparation
\bye